\documentclass[conference]{IEEEtran}
\IEEEoverridecommandlockouts
\usepackage{amsmath,amssymb,amsfonts}
\usepackage{algorithmic}
\usepackage{graphicx}
\usepackage{textcomp}
\usepackage{siunitx}
\usepackage{booktabs}
\usepackage{xcolor}
\usepackage{authblk}

\usepackage[sorting=none]{biblatex}
\def\BibTeX{{\rm B\kern-.05em{\sc i\kern-.025em b}\kern-.08em
    T\kern-.1667em\lower.7ex\hbox{E}\kern-.125emX}}
\bibliography{lit.bib}

\begin{document}
\title{De-Identification of Medical Imaging Data: A Comprehensive Tool for Ensuring Patient Privacy}
\author[*, 1, 2, 5]{Moritz Rempe}
\author[*, 1, 2]{Lukas Heine}
\author[1, 2]{Constantin Seibold}
\author[1, 2]{Fabian Hörst}
\author[1, 3, 4, 5]{Jens Kleesiek}
\affil[*]{Equal contribution}
\affil[1]{Institute for AI in Medicine (IKIM), University Hospital Essen, Girardetstraße 2, 45131 Essen, Germany}
\affil[2]{Cancer Research Center Cologne Essen (CCCE), University Medicine Essen, Hufelandstraße 55, 45147 Essen, Germany}
\affil[3]{RACOON Study Group, Site Essen, Essen Germany }
\affil[4]{Department of Physics of the Technical University Dortmund, Otto-Hahn-Straße 4a, 44227 Dortmund, Germany}
\affil[5]{German Cancer Consortium (DKTK), Partner Site Essen, Hufelandstraße 55, 45147 Essen}
\setcounter{Maxaffil}{0}
\renewcommand\Affilfont{\itshape\small}
\maketitle              % typeset the header of the contribution
\begin{abstract}
% The abstract should briefly summarize the contents of the paper in
% 15--250 words.
Medical data employed in research frequently comprises sensitive patient health information (PHI), which is subject to rigorous legal frameworks such as the General Data Protection Regulation (GDPR) or the Health Insurance Portability and Accountability Act (HIPAA). Consequently, these types of data must be pseudonymized prior to utilisation, which presents a significant challenge for many researchers. Given the vast array of medical data, it is necessary to employ a variety of de-identification techniques. To facilitate the anonymization process for medical imaging data, we have developed an open-source tool that can be used to de-identify DICOM magnetic resonance images, computer tomography images, whole slide images and magnetic resonance twix raw data. Furthermore, the implementation of a neural network enables the removal of text within the images. The proposed tool automates an elaborate anonymization pipeline for multiple types of inputs, reducing the need for additional tools used for de-identification of imaging data. We make our code publicly available at \url{https://github.com/code-lukas/medical_image_deidentification}.
\end{abstract}


\section{Introduction}

Working with real world medical data confers a substantial advantage upon research in comparison to artificially generated data, as it can reflect the actual distribution of data within our world. Nevertheless, using patient data raises the issue of privacy. Medical data - including imaging scans - contain sensitive information which has to be strictly protected \cite{murdoch2021privacy}. Before working with medical data, or distributing it to other researchers, these information have to be de-identified. One example is the RACOON project, which is a consortium of multiple German research sites that exchange clinical data. This motivates the necessity for a uniform deidentification pipeline to ensure that data contributed by individual sites adheres to predefined conventions and is straightforward to implement into existing workflows.

This task can become tedious with different types of imaging data. Clinicians have several medical imaging techniques at their disposal, including magnetic resonance imaging (MRI) and computer tomography (CT), both of which have different ways of storing the resulting data. Some of these storage formats are Digital Imaging and Communications in Medicine (DICOM) \cite{mildenberger2002introduction}, the Neuroimaging Informatics Technology Initiative (NIfTI) \cite{data2004nifti} format, or raw data formats such as Siemens twix (.dat) \cite{larobina2014medical}. Each of these formats has different attributes and information stored, requiring the use of several different tools to anonymize this range of data, leading to potential confusion and errors. 

Most de-identification tools focus on the stored metadata. Medical imaging data contains more sensitive data than just the meta information stored. In particular, brain scans can lead to privacy violations, as they can be subjected to full facial reconstructions if they contain the skull \cite{schwarz2019identification}. With the advent of machine learning, facial reconstruction and general identification of patients based on medical imaging data achieve more accurate results \cite{packhauser2022deep}. Thus, in addition to meta information de-identification, further anonymization steps, such as skull-stripping is needed, which removes the skull from the scan \cite{bischoff2007technique}.

\section{Related Work}

With the rise of digital medical imaging files, the need for anonymization tools arose. Since then, several tools have been proposed and established for different file formats and applications. Most of these anonymization tools perform a specific task, often for only one file format.

Mason, et. al \cite{mason2011t} propose a DICOM metadata anonymization tool using the \textit{pydicom} framework. The \textit{Freesurfer} library \cite{fischl2012freesurfer} contains tools for skull-stripping as well as defacing of MRI NIfTI data. These two tools already show a major problem for practitioners in the medical domain: there is a tool for almost every anonymization task, but not a tool working for all the major file and imaging formats. Additionally, tools like the deface algorithm by \textit{Freesurfer} can be very time consuming, slowing down post-processing pipelines.

When it comes to anonymization tools for WSI data, the literature is rather scarce, especially when looking at open-source software. Bisson et. al. \cite{bisson2023anonymization} present in their work a open-source C-based tool for WSI anonymization which works on a variety of native WSI file formats. Nevertheless it does not work on WSI DICOM files.

\section{Material and Methods}

\subsection{Medical Datatypes}

There are numerous different ways to store the acquired digital data, due to the wide range of medical imaging techniques, such as magnetic resonance imaging (MRI), computer tomography (CT), ultrasound imaging (US) and Whole Slide Images (WSI). The most popular data types are \cite{larobina2014medical}:
\begin{itemize}
 \item Analyze
 \item DICOM
 \item NIfTI
 \item MINC \cite{vincent2016minc}
 \item JP(E)G, PNG
\end{itemize}

While Analyze, MINC and NIfTI are data formats used for research purposes, DICOM (and derivatives, such as DICOM-WSI in pathology) is the file format mainly used in clinical practice. Additionally, most post-processing software is compatible with these two formats, which is not the case for Analyze and MINC.
JP(E)G and PNG are file formats mainly for (clinical) photographic images, for example of wounds or other surface inspections.

In addition to the number of different file format standards, there are raw data formats from different vendors, an example of which is the twix raw data format for MRI data from Siemens.

\subsection{De-identification}

\begin{figure}
  \includegraphics[width=0.45\textwidth]{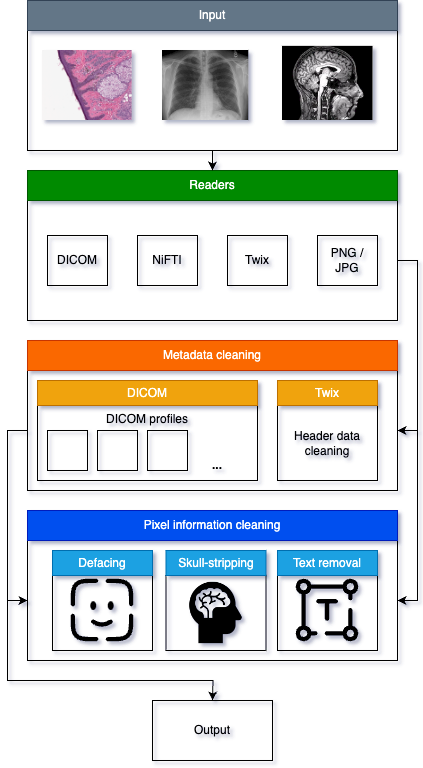}
  \caption{Overview of the proposed de-identification tool. The input data is first read, specific to its data type. Different optional anonymization steps can be performed. Metadata removal or pixel data cleaning, including skull-stripping, defacing or text removal can be performed for all common medical datatypes.}
  \label{fig:overview}
\end{figure}

Medical image anonymization consists of multiple separate tasks:
\begin{itemize}
    \item{Metadata anonymization}
    \item{Defacing}
    \item{Skull-stripping}
    \item{Text removal}
    \item{WSI Anonymization}
\end{itemize}

Metadata anonymization aims to remove all person related meta information in the data header, such as name, sex, date of birth or diagnosis. Because the different data types have different metadata headers, this task spans a wide range of variation in what data needs to be removed.

For DICOM images, the standard defines a guideline on how to handle metadata \cite{dicom2024ps315}. In short, a basic profile on how different metadata information should be handled. At the time of writing, the standard defines ten additional profiles that extend the basic profile and allow limited preservation of some metadata tags if need be. We provide these profiles to be used out of the box to ease the burden of manually checking for conformity. In addition users may add profiles tailored to their specific needs and legislation. 

Defacing is the task of removing the face from a medical image scan, while leaving the rest of the image intact. 
Common tools for this task are \textit{mri\-deface} (\textit{Freesurfer}) \cite{bischoff2007technique} or \textit{pydeface} \cite{gulban2019poldracklab}.
While defacing does not remove the whole skull, the aim of skull-stripping is to only output the brain, thus preserving relevant anatomical information. This approach is more common and thus more researched, with the most popular tools being \textit{SynthStrip} \cite{hoopes2022synthstrip} or the \textit{Brain Extraction Tool} (BET) \cite{smith2002fast}. Isensee et al. utilize deep learning for brain extraction with their \textit{HD-BET} model \cite{isensee2019automated}.
While these tools are well introduced in clinical and research practice, they do come with downsides. One major issue is the long computation times of all approaches. \textit{mri\-deface} and \textit{pydeface} both need more than a minute per volume, making it unfeasible for real-time usage.

The proposed tool is able to anonymize a wide range of data types, including MRI, CT and WSI DICOM files, as well as NIfTI and Siemens MRI raw twix data files (.dat). The DICOM files are anonymized by modifying the metadata according to the predefined ruleset and then performing skull-stripping. The NIfTI file header is discarded before performing skull-stripping. Because the MRI raw data does not display image domain data, but is in the k-Space, this data does not undergo skull-stripping, but its header is anonymized.

An overview of the proposed tool is shown in Fig. \ref{fig:overview}. 

\subsection{Technical requirements}

We provide the proposed tool as a python3 CLI application as well as a standalone docker container, which can be found at https://hub.docker.com/r/morrempe/hold. The technical frameworks used can be found in in Tab. \ref{tab:libraries}.
\begin{table}[!htb]
\centering
\def\arraystretch{1.3}
\caption{Overview of third-party modules}
\begin{tabular}{c|c}
Name & version \\
\hline
pandas & 2.0.1 \\
tqdm & 4.64.1 \\
torch & 2.2.0 \\
timm & 0.9.2 \\
numpy & 1.23.5 \\
torchvision & 0.17.0 \\
PyYAML & 6.0 \\
pydicom & 2.3.1 \\
deid & 0.3.21 \\
torchmetrics & 0.11.4 \\
pathlib & 1.0.1 \\
nibabel & 5.2.1 \\
scipy & 1.13.1 \\
torchio & 0.19.6 \\
tesseract & 5.4.1 \\
\end{tabular}
\label{tab:libraries}
\end{table}

\begin{figure}[!htb]
    \centering
    \includegraphics[width=7cm]{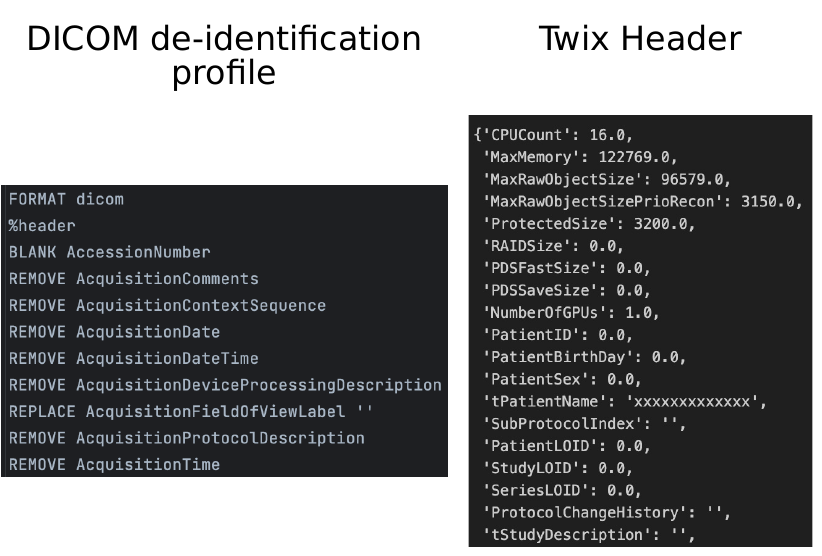}
    \caption{Exemplary excerpt of a DICOM de-identification profile (left) and part of an exemplary anonymized twix header (right). All patient related information are anonymized by either replacing the values with zeros or ’x’.}
    \label{fig:metadata_anonymization}
\end{figure}

\subsection{Datasets}

For the wide variety of different de-identification tasks, multiple different datasets are necessary for training and testing. For training of the skull-stripping approach, we used three publicly available datasets:

\begin{itemize}
    \item{Neurofeedback Skull-stripped (NFBS) repository \cite{eskildsen2012beast}}
    \item{Calgary-Campinas-359 (CC-359) dataset \cite{warfield2004simultaneous}}
    \item{Synthstrip validation dataset \cite{hoopes2022synthstrip}}
\end{itemize}

The NFBS dataset consists of 125 T1 weighted scans, with a resolution of $1\si{\milli\meter}^2$, of subjects between 21 to 45 years old. The ground truth, consisting of skull-strip segmentations, is generated with the STAPLE algorithm \cite{warfield2004simultaneous}, which combines ground truth from different annotators. The CC-359 dataset consists of 354 T1-weighted individual subjects from different scanner manufactures (Siemens, GE \& Philips). The Synthstrip validation dataset is a publicly available subset of the data used to train the Synthstrip dataset and consists of 24 volumes of different sequences and modalities, including MRI, CT and PET. For testing we use the Synthstrip test-dataset. The authors already give a test-split, which includes different modalities, such as diffusion weighted imaging (DWI), T1, T2 and FLAIR, as well as infant subjects. Thus, the total test dataset consists of 558 subjects, with different amounts of slices per volume.

Training of the proposed defacing algorithm was performed on a custom dataset containing 17 volumes of the GBM dataset and 20 volumes of the synthstrip dataset. We chose the volumes such that a big variety of modalities is covered in the dataset. Testing is conducted on 18 subjects of the "IXI-T1" cohort in the synthstrip dataset, as well as 3 infant subjects. The ground truth for the training of the defacing dataset is created by applying the \textit{pydeface} algorithm, as it yielded better results than the \textit{mri\-deface} algorithm.

The proposed text removal algorithm is tested on an internal ultrasound imaging dataset, consisting of 262 2D scans of different anatomies, including kidney and liver.

\section{Experiments}
\subsection{Meta-information de-identification}

Twix data contains multiple headers. While its DICOM header, saved as ['hdr'], can be anonymized like any other DICOM header, the much larger header ['hdr\_string'] is often overlooked, but contains all the same information as the general header, as well as a detailed overview of the scan settings.
If anonymization of twix data is only performed on the DICOM header, all the information contained in the latter will be pasted back into ['hdr'] when saving the ``anonymized'' file. This brings up the necessity and chances of the proposed tool, which scans the hdr\_string for all information which need to be anonymized and replaces them accordingly.

Part of an exemplary anonymized twix header can be seen in Fig. \ref{fig:metadata_anonymization} (right). The patient related metadata is either replaced with zeros, such as "PatientID" and "PatientBirthday", while text-fields, such as "tPatientName" are replaced by "x".

\subsection{Skull-stripping \& Defacing}

For skull-stripping and defacing a 3D MedNext \cite{roy2023mednext} is used, which is not pretrained.

Each input NIfTI file is first reoriented into the ('R', 'A', 'S') orientation, first dimension pointing to the right hand side of the head, second dimension towards the Anterior aspect of the head and the third dimension towards the top of the head. The data is then converted into pytorch tensors. In case of DICOM input data, the data is first converted into NIfTI files via the \textit{pydicom} library \cite{mason2011t}.
For training, we used augmentations, including randomly inserted noise or spiking, as well as deformations, to provide our models with the ability to also work with data which might contain artifacts.

All augmentations are implemented with torchIO \cite{perez2021torchio}, a python library specialized on preprocessing of medical images. The initial learning rate is set to $5e^{-4}$ with a cosine annealing schedule and a DICE loss function. All input data is normalized and standardized. A batch size of one is used and the input is resized to a shape of [64, 224, 224]. Training was performed on a single NVIDIA A6000 GPU with 48GB of graphics memory.

\begin{figure*}[!htb]
  \centering
  \includegraphics[width=17cm]{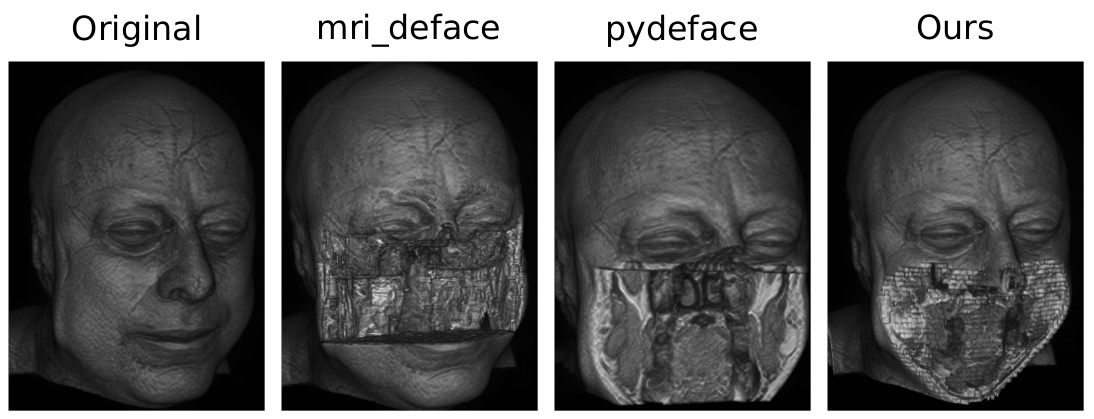}
  \caption{Comparison of the defacing results of different defacing algorithms. While the result of \textit{pydeface} and the proposed algorithm are similar, \textit{pydeface} additionally cuts off the shoulder region of the scan, while taking 260 times longer on average than the proposed algorithm.}
  \label{fig:deface}
\end{figure*}

To compare our defacing algorithm with other existing state-of-the-art algorithms, we developed a "defacing-score". The output data of the deface algorithm is first loaded into the medical image viewer \textit{ITK-SNAP} \cite{py06nimg} to generate a volumetric view of the scan. The front facing view of this volume is then saved as an image. An exemplary output of our defacing model can be seen in Fig. \ref{fig:deface}. By applying the face recognition model by A. Geitgey \cite{geitgey2019face}, we can then distinguish between correct and flawed defacing results. The defacing score is calculated by the amount of defaced scans, which the face recognition model cannot classify as faces anymore. 

For the comparison of the skull-stripping algorithms we use the commonly used DICE score. As computation time is crucial for real time algorithms, we additionally track the computation time per volume for each algorithm.

\subsection{Text removal}

\begin{figure*}[!htb]
  \centering
  \includegraphics[width=\textwidth]{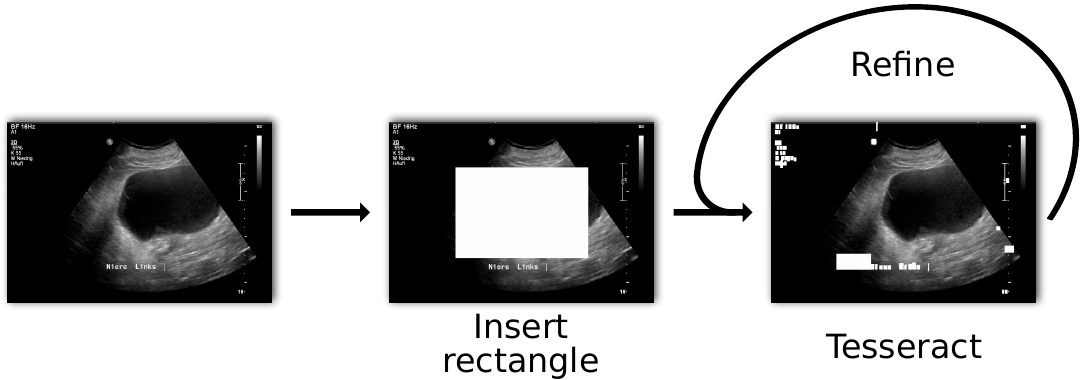}
  \caption{Proposed text removal pipeline at the example of a ultrasound image. By first inserting a rectangle in the center of the scan, tesseract focuses on the text on the side of the image. Possible texts in the middle of the image is then removed in further iterations.}
  \label{fig:text_removal}
\end{figure*}

By utilizing the \textit{tesseract} \cite{kay2007tesseract} algorithm for text detection we propose a novel pipeline for text removal in medical images. To reach consistent anonymization results, the proposed method first inserts a white rectangle in the center of the input image. That way \textit{tesseract} focuses on the border of the scans which holds the majority of the text information of medical scans. In a second iteration, the rectangle is removed and the detection algorithm is applied another time on the already processed image. for the final output, all detected boundary boxes containing letters are then filled and thus covered. 

The proposed text removal pipeline can be seen in Fig. \ref{fig:text_removal}. The method is tested on 262 2D ultrasound scans.

\subsection{DICOM-WSI Anonymization}

DICOM-WSI data is saved in a hierarchical structure, consisting of the different zoom-stages of the image, an overview image and a label image. The label image contains information to identify the slide and patient. We remove these information by overwriting the whole pixel array with zeros. Often the overview image also contains some kind of label which needs to be deidentified, which can be removed by overwriting the left side of the overview image. Additionally to the image information, each DICOM file contains metadata which has to be removed. 

In case of converted DICOM files from other data types, we remove the label file entirely, as the pixel array can not be loaded reliably via pydicom.

\section{Results}

Both proposed algorithms for defacing and skull-stripping perform similar to the compared state-of-the-art algorithms, while significantly reducing the computation time. The proposed defacing algorithm is able to reduce the computation time per volume by up to 260 times, to an average time of $0.88\si{\second} \pm 0.15\si{\second}$, compared to $233.57\si{\second} \pm 59.87\si{\second}$ needed by \textit{pydeface}. Even though \textit{mri-deface} can reduce the computation time per volume to $95.59\si{\second} \pm 18.00\si{\second}$, it still takes on average 105 times longer than the proposed method, while achieving a worse defacing score of $74.38\% \pm 17.24\%$ in comparison to $80.62\% \pm 13.95\%$. The calculation time of the proposed method benefits from GPU-support, but also reaches better times on CPU-only devices. A full overview of the results can be seen in Tab. \ref{tab:1}.
While \textit{mri-deface} struggles with head scans of infants, \textit{pydeface} and the proposed algorithm achieve sound results also on these type of scans. The \textit{pydeface} algorithm removes the shoulder region of most of the scans, while the proposed method does not remove any additional body regions to the face. 

\begin{table}[!htb]
\caption{Defacing scores and computation times per volume for different defacing algorithms. The computation time is given per average volume. While \textit{mri\-deface} and \textit{pydeface} can only be used on CPUs, the proposed method can be used on GPU devices.}
\centering
\def\arraystretch{1.6}
\begin{tabular}{c|c|c}
\hline
{Method} & {Defacing score ($\%) \uparrow$} & {Computation time (s) $\downarrow$}\\
\hline
mri\-deface (CPU) \cite{bischoff2007technique} & 74.38 ± 17.24 & 95.59 ± 18.00 ($\times$108.6)\\
pydeface (CPU) \cite{gulban2019poldracklab} & 76.43 ± 14.53 & 233.57 ± 59.87 ($\times$265.4)\\
Ours (CPU) & \textbf{80.62 ± 13.95} & \textbf{25.07 ± 3.14} ($\times$28.5)\\
Ours (GPU) & \textbf{80.62 ± 13.95} & \textbf{0.88 ± 0.15} \\
\hline
\end{tabular}
\label{tab:1}
\end{table}

Tab. \ref{tab:2} shows the results of different skull-stripping algorithms in comparison to the proposed tool. While the proposed method outperforms the commonly used algorithm BET, it reaches similar results to \textit{Synthstrip} and \textit{HD-BET}. Nevertheless, the proposed method again outperforms both methods in regards of computation times, being 20 times faster than \textit{Synthstrip}, with $0.77\si{\second} \pm 0.19\si{\second}$ against $15.34\si{\second} \pm 2.62\si{\second}$ per volume. 

\begin{table}[!htb]
\caption{DICE score and computation times per volume for different skull-stripping algorithms. The computation time is given per average volume. While BET can only be used on CPU, \textit{Synthstrip}, \textit{HD-BET} and the proposed method offer GPU support.}
\centering
\def\arraystretch{1.6}
\begin{tabular}{c|c|c}
\hline
{Method} & {DSC ($\%) \uparrow$}& {Computation time (s) $\downarrow$}\\
\hline
BET (CPU) & 84.95 ± 14.22 & 4.15 ± 0.79 ($\times$5.4) \\
Synthstrip (GPU) & 96.85 ± 1.03 & 15.34 ± 2.62 ($\times$19.9) \\
HD-BET (GPU) & 95.10 ± 2.87 & 24.26 ± 5.90 ($\times$31.5) \\
Ours (GPU) & 94.00 ± 3.61 & \textbf{0.77 ± 0.19} \\
\hline
\end{tabular}
\label{tab:2}
\end{table}

\begin{figure*}[!htb]
  \centering
  \includegraphics[width=17cm]{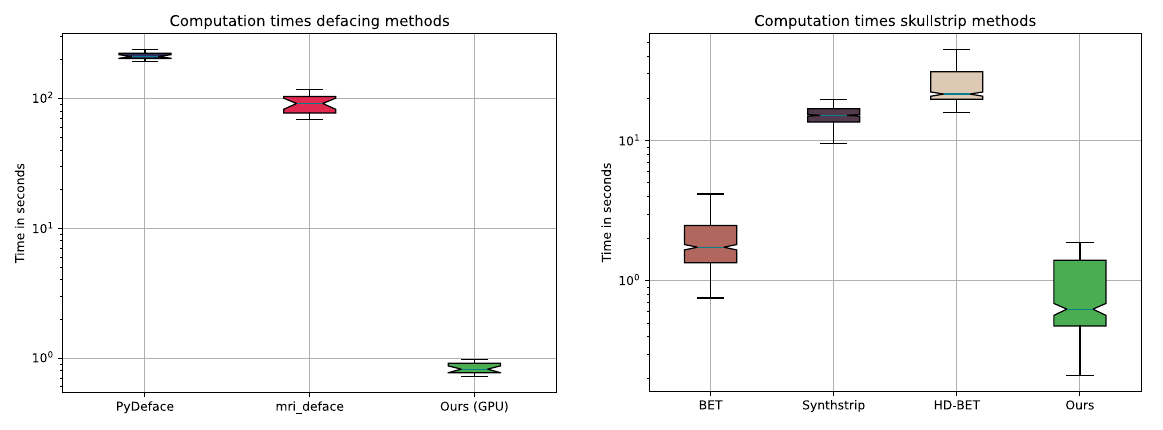}
  \caption{Computation time for defacing and skull-stripping of the compared methods on the Synthstrip test-dataset. The proposed method is faster than the compared state-of-the art algorithms. The y-axis is scaled logarithmically for better visibility. The bold lines inside the plots depict the median value.}
  \label{fig:computation_times}
\end{figure*}

Fig. \ref{fig:computation_times} shows the computation time advantage of the proposed tool. For better visibility we used a logarithmic scale.

\begin{figure*}[!htb]
  \centering
  \includegraphics[width=17cm]{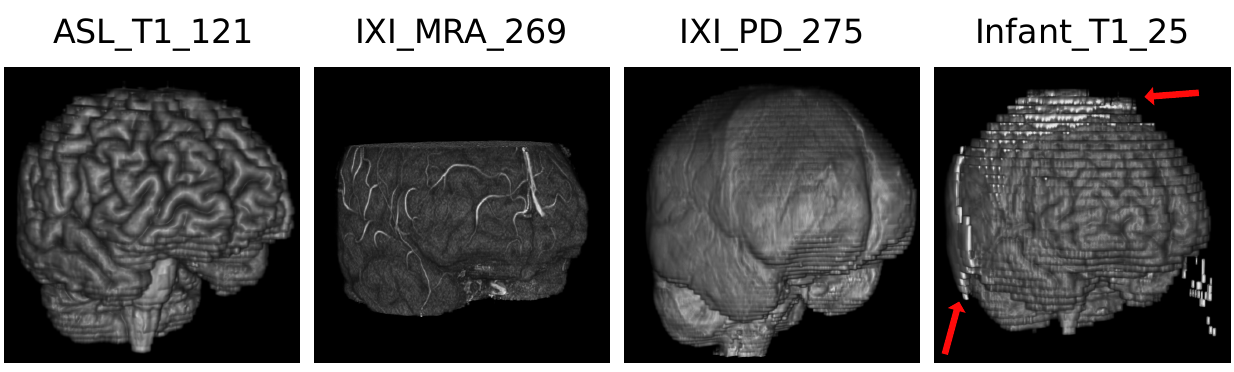}
  \caption{Skull-stripping results of the proposed algorithm. Shown are examples from the \textit{Synthstrip} test dataset, including T1, magnetic resonance angiography (MRA), proton density (PD) and infant T1 scans. The proposed methods produces sound results, but might struggle with parts of infant brains.}
  \label{fig:skullstrip}
\end{figure*}

Exemplary results of the proposed skull-strip algorithm is shown in Fig. \ref{fig:skullstrip}. The test dataset covers a wide range of modalities, including T1, magnetic resonance angiography (MRA), proton density (PD) and infant brain scans. The proposed method is able to perform the task of skull-stripping on all of these different modalities, but struggles with infant brain scans, leaving some skull residues in the output. 

The proposed text removal method reaches sound results, with an de-identification score of $83.59\%$ on the tested ultrasound images. The scan is only counted as deidentified, if all the text present in the image is removed. An example is shown in Fig. \ref{fig:text_removal}. 
% In some scans the proposed methods also inserts boundary boxes inside the tomography scan, which are not actual texts. These faulty removals have no impact on the quality of the scan, as they are mainly spots of noise in the image. 

\section{Discussion}

We present a de-identification tool, which takes on the challenge of combining multiple anonymization steps in one tool, making it easy for clinicians and researchers to de-identify a wide variety of medical data, without the need for different tools. By focusing on better computation times, we are able to speed up the process of deidentifcation, enabling real-time data processing, while achieving state-of-the-art (SOA) results.

Nevertheless, our tool comes with limitations. One being the limited amount of training data and preprocessing steps. While the current SOA algorithms often perform an extensive preprocessing before defacing or skull-stripping, we almost completely do without these steps, with the goal of reducing the computation time. While we are able to reduce the computation times by up to 20 times in skull-stripping with GPU support, we sacrifice minor accuracy gains for the sake of computation time. This speedup is particularly important in clinical practice, when large amounts of data have to be processed on a daily basis.

Additionally, the currently used text detection algorithm removes all text present in the data and does not differentiate between text information. While the use of \textit{tesseract} is common practice, there are more advanced text detection algorithms which could be implemented in future version of the proposed tool, such as the \textit{Document understanding transformer} (Donut) by Kim et al. \cite{kim2022ocr}. 

The proposed WSI anonymization tool is currently only supporting DICOM WSI files. This will be extended to further data types in the future.

\section{Conclusion}

In this work we propose a novel de-identification tool which combines metadata anonymization, defacing, skull-stripping and text removal in one framework. We give researchers and clinicians a tool which focuses on computation time and ease of use to enable them to perform real time de-identification on a wide range of medical data. 
While achieving computation time speedups of up to 260 times in comparison to commonly used tools, we achieve state-of-the-art results with a simple CLI tool, that can be integrated into existing data processing pipelines.
The proposed tool makes the tedious task of installing multiple different de-identification tools superfluous.

\section{Acknowledgement}

This work received funding from the RACOON network under NUM 2.0 [01KX2021], the Bruno \& Helene Jöster Foundation and KITE (Plattform für KI-Translation Essen) from the REACT-EU initiative (https://kite.ikim.nrw/). This work was also supported by DFG RTG 2535 - AI for personalized medicine.

\section{References}

\printbibliography[heading=none]

\end{document}